\input{aipcheck}

\documentclass[
    ,final            
  ]
  {aipproc}

\layoutstyle{8x11single}

\newcommand{\ba}{\begin{eqnarray}}
\newcommand{\ea}{\end{eqnarray}}
\newcommand{\ACal}{{\cal{A}}}

\newcommand{\C}{{\cal {C}}}
\newcommand{\DD}{{\cal {D}}}

\newcommand{\RR}{{}^3{\cal{R}}}

\newcommand{\T}{{\cal{T}}^{(3)}}

\newcommand{\EE}{{\cal{E}}}
\newcommand{\FF}{{\cal{F}}}
\newcommand{\JJ}{{\cal{J}}}
\newcommand{\VV}{{\cal{V}}}

\newcommand{\PP}{{\cal{P}}}
\newcommand{\MM}{{\cal{M}}}
\newcommand{\QQ}{{\cal{Q}}}

\newcommand{\rhoav}{\langle\rho\rangle}

\newcommand{\Thetaav}{\langle\Theta\rangle}

\newcommand{\RRav}{\langle\RR\rangle}
\newcommand{\Aav}{\langle A\rangle}
\newcommand{\Bav}{\langle B\rangle}

\newcommand{\dd}{{\rm{d}}}

\begin{document}

\title[Quasi--local variables and scalar averaging in LTB dust models.]{Quasi--local variables and scalar averaging in LTB dust models.}


\author{Roberto A. Sussman}{
  address={Instituto de Ciencias Nucleares, UNAM, M\'exico D.F. 04510, M\'exico}
}



\begin{abstract}
We introduce quasi--local (QL) scalar variables in spherically symmetric LTB models. If the QL scalars are defined as functionals, they become weighed averages that generalize the standard proper volume averages on space slices orthogonal to the 4--velocity. We examine the connection between QL functions and functionals and the ``back--reaction'' term $\QQ$ in the context of Buchert's scalar averaging formalism. With the help of the QL scalars we provide rigorous proof that back--reaction is positive for (i) all LTB models with negative and asymptotically negative spatial curvature, and (ii) models with positive curvature decaying to zero asymptotically in the radial direction. We show by means of qualitative, but robust, arguments that generic LTB models exist, either with clump or void profiles, for which an ``effective'' acceleration associated with Buchert's formalism can mimic the effects of dark energy.          
\end{abstract}

\maketitle


\section{Introduction.}

  Recent observations apparently reveal that the universe is spatially flat and is undergoing an accelerated expansion. To account for these observations, a large variety of theoretical and empiric models have molded a dominant theoretical paradigm: the ``concordance'' model, based on the assumption that cosmic dynamics appears to be dominated by an elusive source (``dark energy'') that behaves as a cosmological constant or as a fluid with negative pressure (see \cite{review} for a review). 
  
The concordance model also assumes that inhomogeneities play a minimal role in cosmic dynamics at scales over 100 Mpc, and thus can be adequately dealt with in terms of linear perturbations on a FLRW background. This assumption, and the existence of dark energy, has been challenged from various angles~\cite{celerier}. In trying to account with supernovae observations, numerous articles (see \cite{celerier} for a review, see also \cite{InhObs}) show that cosmic acceleration can be reproduced simply by considering large scale inhomogeneities in photon trajectories within the so--called ``homogeneity scale'' (100-300 Mpc). From a theoretical point of view, it has been argued that inhomogeneity implies that observations from distant high redshift sources must be understood in terms of averaged quantities \cite{InhObs,buchert,ave_review,zala,colpel,wiltshire2}, which in homogeneous conditions would be trivially identical with local quantities. Thus, non--linear spatial gradients of the Hubble expansion scalar and quasi--local effective energies would have an important effect in the interpretation of these observations \cite{wiltshire2}.

The spherically symmetric LTB dust models \cite{kras,ltbstuff,suss02} have often been used to test the effects of inhomogeneity in cosmic observations, as well as the issue of back--reaction in the context of Buchert's spatial averaging~\cite{LTBave1,LTBave2}. In the present article we examine the dynamical equations of there models in terms of suitably defined quasi--local variables and of spatially averaged scalars~\cite{sussQLnum, sussQL}. We show that the back--reaction term is the difference between squared fluctuations of the expansion Hubble scalar, which in turn are related to spatial gradients of its average and quasi--local equivalent. Necessary conditions for a positive ``effective'' accelerations, mimicking the effect of dark energy, follow from comparing these fluctuations, and are satisfied as long as spatial curvature is negative in a sufficiently large averaging domain (even if smaller domains contain bound structures). Since these conditions are compatible with the observed void dominated structure, it is highly likely that such ``effective'' acceleration could be observed. However, further steps in this direction require a more elaborate numerical study of these models (see \cite{sussBR}).     
  
\section{LTB dust models in the ``fluid flow'' description.}

Spherically symmetric inhomogeneous dust sources are usually described by the well known Lema\^\i tre--Tolman--Bondi metric and energy--momentum tensor in a comoving frame~\cite{kras,ltbstuff,suss02}
\ba\dd s^2 &=& -c^2\dd t^2+ \frac{R'{}^2}{\FF^2}\dd r^2+R^2\left(\dd\theta^2+\sin^2\theta\dd\phi^2\right).\label{ltb}\\
T^{ab} &=& \rho\,u^au^b, \label{Tab}\ea
where $R=R(ct,r)$,\, $R'=\partial R/\partial r,\,\FF=\FF(r),\, u^a=\delta^a_0$ and $\rho(t,r)$ is the rest matter--energy density. The field equations  $G^{ab}=\kappa T^{ab}$ (with $\kappa=8\pi G/c^2$) for (\ref{ltb}) and (\ref{Tab}) reduce to
\ba \dot R^2 &=& \frac{2M}{R}+\FF^2-1,\label{eqR2t}\\
2M' &=& \kappa\,\rho\,R^2R',\label{eqrho1}\ea
where $M=M(r)$ and $\dot R=u^a\nabla_a R$. The sign of $\FF^2-1$ determines the zeroes of $\dot R$ and thus classifies LTB models in terms of the following kinematic classes:
\ba \FF = 1,&{}&\qquad \hbox{parabolic models}\nonumber\\
\FF \geq 1, &{}& \qquad \hbox{hyperbolic models}\nonumber\\
0 <\FF\leq 1, &{}&\qquad \hbox{"open" elliptic models}\nonumber\\
-1\leq \FF\leq 1, &{}&\qquad \hbox{"closed" elliptic models}\nonumber\\\ea
where by ``open'' or ``closed'' models we mean the cases where the $\T$ are, respectively, topologically equivalent to $\textrm{\bf {R}}^3$ and  $\textrm{\bf {S}}^3$. Regularity conditions \cite{ltbstuff,suss02} require $\FF'\geq 0$ for all $r$ in hyperbolic models, while in open elliptic models $\FF'$ can be, either negative for all $r>0$, or it can have a zero $\FF'(y)=0$, so that $\FF'\leq 0$ for $0\leq x\leq y$ and $\FF'> 0$ for $x> y$. 

Besides $\rho$ and $R$ given above, other covariant objects of LTB spacetimes are the expansion scalar $\Theta$, the Ricci scalar $\RR$ of the hypersurfaces $\T$ orthogonal to $u^a$, the shear tensor $\sigma_{ab}$ and the electric Weyl tensor $E^{ab}$: 
\ba \Theta &=& \tilde\nabla_au^a=\frac{2\dot R}{R}+\frac{\dot R'}{R'},\qquad
\RR = \frac{2[(1-\FF)\,R]'}{R^2R'},\label{ThetaRR}\\
\sigma_{ab} &=& \tilde\nabla_{(a}u_{b)}-(\Theta/3)h_{ab}=\Sigma\,\Xi^{ab},\qquad
E^{ab}=  u_cu_d C^{abcd}=\EE\,\Xi^{ab},\label{SigEE}\ea
where $h_{ab}=u_au_b-g_{ab}$,\, $\tilde\nabla_a = h_a^b\nabla_b$,\, and $C^{abcd}$ is the Weyl tensor, $\Xi^{ab}=h^{ab}-3\eta^a\eta^b$ with $\eta^a=\sqrt{h^{rr}}\delta^a_r$ being the unit radial vector orthogonal to $u^a$.  The scalars $\EE$ and $\Sigma$ in (\ref{SigEE}) are
\begin{equation}\Sigma = \frac{1}{3}\left[\frac{\dot R}{R}-\frac{\dot R'}{R'}\right],\qquad
\EE = -\frac{\kappa}{6}\,\rho+ \frac{M}{R^3}.\label{SigEE1}\end{equation}
The dynamics of LTB spacetimes can be fully characterized by the local covariant scalars $\{\mu,\,\Theta,\,\Sigma,\,\EE,\,\RR\}$. Given the covariant ``1+3'' slicing afforded by $u^a$,  the evolution of the models can be completely determined by a ``fluid flow''  description of scalar evolution equations for these scalars (as in \cite{1plus3}):
\ba
\dot\Theta &=&-\frac{\Theta^2}{3}
-\frac{\kappa}{2}\,\rho-6\,\Sigma^2,\label{ev_theta_13}\\
\dot \rho &=& -\rho\,\Theta,\label{ev_mu_13}\\
\dot\Sigma &=& -\frac{2\Theta}{3}\,\Sigma+\Sigma^2-\EE,
\label{ev_Sigma_13}\\
 \dot\EE &=& -\frac{\kappa}{2}\rho\,\Sigma
-3\,\EE\,\left(\frac{\Theta}{3}+\Sigma\right),\label{ev_EE_13}\ea
together with the spacelike constraints  
\begin{equation} 
\left(\Sigma+\frac{\Theta}{3}\right)'+3\,\Sigma\,\frac{R'}{R}=0,\qquad
\frac{\kappa}{6}\rho'
+\EE\,'+3\,\EE\,\frac{R'}{R}=0,\label{cSE_13}\end{equation}
and the Friedman equation (or ``Hamiltonian'' constraint)
\begin{equation}\left(\frac{\Theta}{3}\right)^2 = \frac{\kappa}{3}\, \mu
-\frac{\RR}{6}+\Sigma^2,\label{cHam_13}\end{equation}
The solutions of system
(\ref{ev_theta_13})--(\ref{cHam_13}) are equivalent to the solution of the field plus conservation equations $\nabla_bT^{ab}=0$.

\section{Proper volume average: Buchert's formalism.}

The time slicing defined by $u^a$ defines as the space slices the hypersurfaces $\T$ marked by constant $t$, whose metric is $h_{ab}= u_au_b+g_{ab}$ and their proper volume element is 
\begin{equation} \dd\VV_p = \sqrt{\textrm{det}(h_{ab})}\,\dd r\,\dd\theta\,\dd\phi=\FF^{-1}\,R^2\,R'\,\sin\theta\,\dd r\,\dd\theta\,\dd\phi.\label{propvol}\end{equation}
Consider spherical comoving regions of the form
\begin{equation} \DD[r]=\vartheta[r]\times\textrm{\bf{S}}^2\subset \T,\qquad \vartheta[r]\equiv\{x\in {\textrm{\bf{R}}}\,|\,0\leq x\leq r\}\label{domain}\end{equation}
where ${\textrm{\bf {S}}}^2$ is the unit 2--sphere and $x=0$ marks a symmetry center. We introduce now the following definitions:

\begin{quote}

\noindent Let $X(\DD[r])$ be the set of all smooth integrable scalar functions in $\DD$, then\\

\noindent 
 {\underline{Definition 1.}}  For every $A\in X(\DD[r])$, the p-map is defined as
\begin{equation}\JJ_p:X(\DD[r])\to X(\DD[r]),\qquad A_p=\JJ_p(A)=\frac{\int_{\DD[r]}{A\,\dd\VV_p}}{\int_{\DD[r]}{\dd\VV_p}}=\frac{\int_0^r{A \FF^{-1}R^2 R'\dd x}}{\int_0^r{\FF^{-1} R^2 R'\dd x}}.\label{pmap}\end{equation}
where $\int_0^r{..\,\dd x}=\int_{x=0}^{x=r}{..\,\dd x}$. The scalar functions $A_p:\DD\to {\bf\rm{R}}$ that are images of $\JJ_p$ will be denoted by ``p--functions''. In particular, we will call $A_p$ the p--dual of $A$. \\

\noindent 
 {\underline{Definition 2.}} For every $A\in X(\DD)$, the proper volume average is the functional
\begin{equation}\langle \, \,\, \rangle_p[r]:X(\DD)\to \textrm{\bf{R}},\qquad \Aav_p[r]=A_p(r).\label{pavemap}\end{equation}
The real number $\Aav_p[r]$, associated to the full domain $\DD[r]$, will be denoted by the proper volume average of $A$ on $\DD[r]$. In order to simplify notation, we will drop the ``$[r]$'' and ${}_p$ symbols and express (\ref{pavemap}) simply as $\Aav$, as it is clear that the average is evaluated as a proper volume integral and it depends on the domain boundary as a functional.

\end{quote} 

\noindent
The proper volume average (as well as the p--functions) satisfy the following commutation rule:
\begin{equation} 
\Aav\,\dot{}-\langle\dot A\rangle = \langle \Theta A\rangle-\Thetaav\Aav,\label{conmr1}
\end{equation}
It is important to emphasize that only the functionals $\Aav$ can be considered as average distributions, as they satisfy $\Aav\rangle = \Aav$, and thus 
\ba 
\langle (A-\Aav)^2\rangle &=& \langle A^2\rangle-\Aav^2,\label{var}\\
\langle (A-\Aav)\,(B-\Bav)\rangle &=& \langle AB\rangle-\Aav\Bav,\label{cov}\ea
which define variance and covariance moment definitions for continuous random variables. It is straightforward to verify that the functions $A_p$ do not satisfy these equations. The relation between the p--functions $A_p$ and the average $\Aav$ is illustrated by figure 1.
\begin{figure}[htbp]
\includegraphics[width=5in]{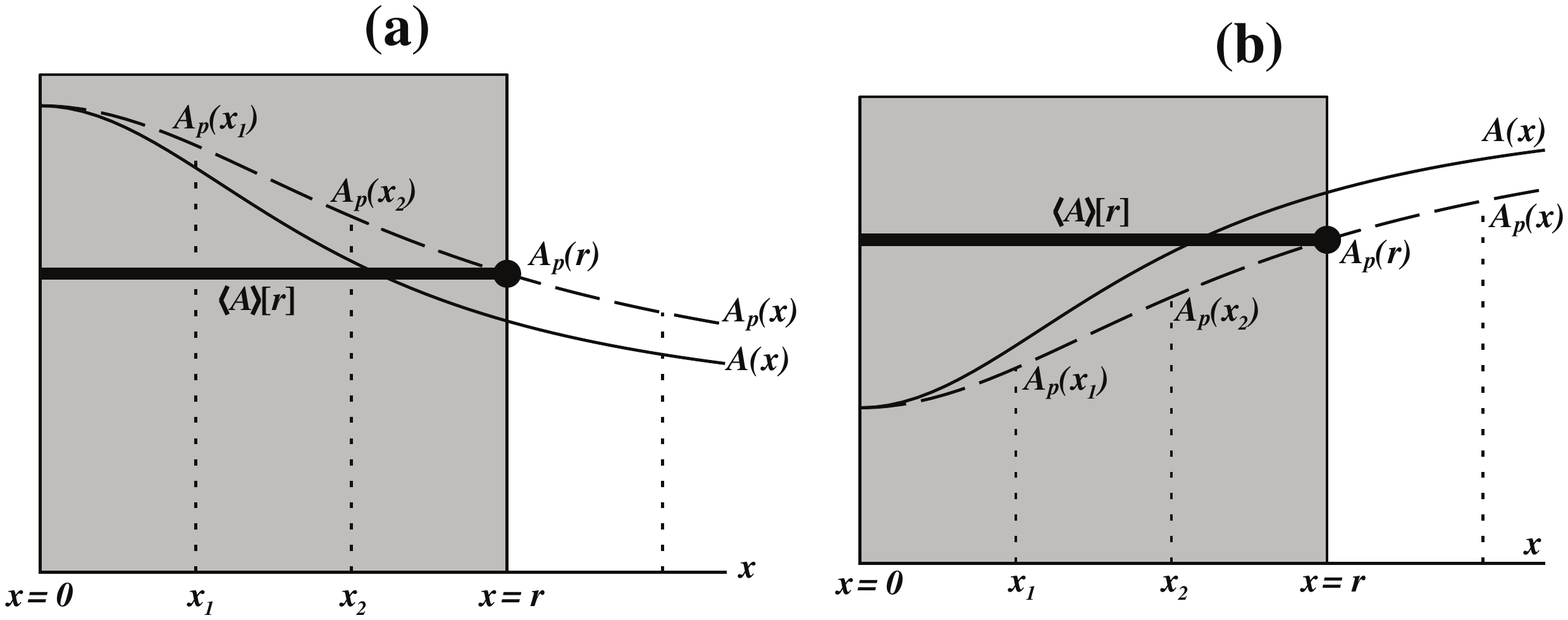}
\caption{{\bf The difference between $A_p$ and $\Aav$.} The figure displays the radial profile of a scalar function $A(x)$ (solid curve) along a regular hypersurface $\T(t)$,  together with its dual p--function $A_p(x)$ (dotted curve) defined by (\ref{pmap}). Panels (a) and (b) respectively display the cases when $A'\leq 0$ (``clump'') and $A'\geq 0$ (``void''). The average functional (\ref{pavemap}) assigns the real number $\Aav[r]$ to the full domain (shaded area) marked by $\vartheta[r]=\{x\,|\, 0\leq x\leq r\}$, whereas the function $A_p$ varies along this domain. Hence, $A_p$ and $\Aav$ are only equal at the domain boundary $x=r$, and so they satisfy the same differentiation rules locally, {\it i.e.} $\dot A_p(r)=\Aav\,\dot{}[r]$ and $A_p'(r)=\Aav'[r]$, but behave differently when integrated along the domain. Notice that, from (\ref{prop1}) and (\ref{prop2}), if $A'\leq 0$ in all $\vartheta[r]$ then $A-\Aav\leq 0$ and the opposite situation occurs if $A'\geq 0$. This figure also applies for the quasi--local functions and averages.}
\label{fig1}
\end{figure}

The well known evolution equations of Buchert's formalism \cite{buchert} follow by applying the proper volume average functional (\ref{pavemap}) to both sides of the energy balance (\ref{ev_mu_13}), Raychaudhuri (\ref{ev_theta_13})  and Friedman (\ref{cHam_13}) equations, and then using (\ref{conmr1}) and (\ref{var})--(\ref{cov}) to eliminate averages $\langle \dot A\rangle$ in terms of derivatives of averages $\Aav\,\dot{}$ and squares of averages as averages of squares. The averaged forms of these  equations are:
\ba \langle\dot\rho+\rho\Theta\rangle &=& \rhoav\,\dot{}+\rhoav\Thetaav=0,\label{ave_ev_rho}\\ 
\Thetaav\,\dot{}+ \frac{\Thetaav^2}{3}&=&-\frac{\kappa}{2}\rhoav+\QQ,\label{ave_Raych}\\
 \frac{\Thetaav^2}{9}-\frac{\kappa}{3}\,\rhoav &=&-\frac{\RRav+\QQ}{6},\label{ave_Fried}\ea
where the kinematic ``back--reaction'' term, $\QQ$, is given by
\begin{equation}\QQ[r] \equiv \frac{2}{3}\,\langle(\Theta-\Thetaav)^2\rangle-6\,\langle\Sigma^2\rangle.\label{QQ}\end{equation}
Equation (\ref{ave_ev_rho}) simply expresses the compatibility between the averaging (\ref{pavemap}) and the conservation of rest mass, but (\ref{ave_Raych}) and (\ref{ave_Fried}) lead to an interesting re--interpretation of the dynamics because of the presence of $\QQ$. This follows by re--writing these equations as 
 \label{ave_eveqs_ef}\ba \Thetaav\,\dot{}+ \frac{\Thetaav^2}{3}&=&-\frac{\kappa}{2}\left[\rho_{\rm{eff}}+3\,P_{\rm{eff}}\right],\label{ave_eveqs_ef1}\\
 \frac{\Thetaav^2}{9} &=& \frac{\kappa}{3}\,\rho_{\rm{eff}},\label{ave_eveqs_ef2}\ea     
where the ``effective'' density and pressure are
\begin{equation} \kappa\,\rho_{\rm{eff}} \equiv \kappa\,\rhoav -\frac{\RRav+\QQ}{2},\qquad
\kappa\,P_{\rm{eff}} \equiv \frac{\RRav}{6}-\frac{\QQ}{2}, \label{rhoPefe}\end{equation}     
The compatibility condition between (\ref{ave_Raych}), (\ref{ave_Fried}) and (\ref{QQ})  is given by the following relation between $\dot\QQ$ and $\RRav\,\dot{}$ (as in \cite{buchert,LTBave2}):
\begin{equation} \dot\QQ +2\Thetaav\,\QQ+\frac{2}{3}\Thetaav\,\RRav+\RRav\,\dot{}=0,\label{consistQR}\end{equation}
which is equivalent to the compatibility between the time derivative of (\ref{cHam_13}), equations (\ref{ev_theta_13})--(\ref{ev_mu_13}), the commutation rule (\ref{conmr1}) and the variance (\ref{var}) for $\Theta$ and $\RR$. From (\ref{ave_eveqs_ef1}), the condition for an ``effective'' cosmic acceleration mimicking dark energy is 
\begin{equation}\frac{\kappa}{2}\,\left[\rho_{\rm{eff}}+3\,P_{\rm{eff}}\right]=\frac{\kappa}{2}\,\rhoav-\QQ < 0,\label{efeacc}\end{equation}
which, apparently, could be possible to fulfill for a sufficiently large and positive back reaction $\QQ$. We will evaluate this condition for spherically symmetric LTB dust solutions (see \cite{sussBR} and \cite{LTBave1,LTBave2} for previous work on this). 

\section{Quasi--local (QL) variables.}

The Misner--Sharp quasi--local mass--energy function, $\MM$, is a well known invariant in spherically symmetric spacetimes~\cite{MSQLM,hayward1}. For LTB dust models  (\ref{ltb})--(\ref{Tab}) it satisfies the equations
\begin{equation} 2\MM'=\kappa \rho\,R^2R',\qquad 
2\dot\MM = 0,\qquad\Rightarrow \MM = M(r)\label{Mr}\end{equation}
where $M$ is the function appearing in the field equation (\ref{eqrho1}). Comparing (\ref{Mr}) and (\ref{eqrho1}) suggest obtaining an integral expression for $\MM$ that can be related to $R$ and $\dot R$. This integral along the $\T$ exists and is bounded if we consider an integration domain of the form (\ref{domain}) containing a symmetry center \cite{hayward1}. Since $\MM(ct,0)=0$ for all $t$, we integrate both sides of (\ref{eqrho1}) and also (\ref{Mr}). This allows us to define a scalar $\rho_q$ as
\begin{equation} \frac{\kappa}{3}\rho_q\equiv \frac{2\MM}{R^3} = \frac{2M}{R^3}= \frac{\kappa}{3}\, \frac{\int_0^r{\rho R^2 R'\dd x}}{\int_0^r{R^2 R'\dd x}},\label{QLrho}\end{equation}
This integral definition of $\rho_q$, which is related to $\rho$ and to the quasi--local mass--energy function, $\MM$, motivates us to generalize it to other scalars by means of the following:\\

\begin{quote}

\noindent 
 {\underline{Definition 3: Quasi--local (QL) scalar map}}. Let $X(\DD)$ be the set of all smooth integrable scalar functions in $\DD$. For every $A\in X(\DD)$, the quasi--local map is defined as
\begin{equation}\JJ_q:X(\DD)\to X(\DD),\qquad A_q=\JJ_q(A)=\frac{\int_{\DD[r]}{A\,\FF\,\dd\VV_p}}{\int_{\DD[r]}{\FF\,\dd\VV_p}}=\frac{\int_0^r{A R^2 R'\dd x}}{\int_0^r{R^2 R'\dd x}}.\label{QLmap}\end{equation}
The scalar functions $A_q:\DD\to {\bf\rm{R}}$ that are images of $\JJ_q$ will be denoted by ``quasi--local'' (QL) scalars. In particular, we will call $A_q$ the QL dual of $A$. Notice that a quasi--local average can also be defined by means of a functional with the correspondence rule (\ref{QLmap}), but we will not need it in this article. 

\end{quote}

\noindent
Applying the map (\ref{QLmap}) to the scalars $\Theta$ and $\RR$ in (\ref{ThetaRR}) we obtain
\begin{equation} \Theta_q =\frac{3\dot R}{R},\qquad \RR_q =\frac{6(1-\FF^2)}{R^2}.\label{QL_TR}\end{equation}
Applying now (\ref{QLmap}) to $\rho$, comparing with (\ref{eqR2t}) and (\ref{eqrho1}), and using (\ref{QL_TR}), these two field equations yield 
\ba \left(\frac{\Theta_q}{3}\right)^2 &=& \frac{\kappa}{3}\rho_q -\frac{\RR_q}{6},\label{QLfried}\\
\dot \Theta_q &=&  -\frac{\Theta_q^2}{3}-\frac{\kappa}{2}\rho_q.\label{QLraych}\\
\dot\rho_q &=& -\rho_q\,\Theta_q.\label{QLebal}\ea
which are identical to the Friedman, Raychaudhuri and energy balance equations for dust FLRW cosmologies, but given among QL scalars (notice that we are using here the QL functions, not QL averages). By applying (\ref{QLmap}) to (\ref{SigEE1}) the remaining covariant scalars $\{\Sigma,\,\EE\}$ can be expressed as deviations or fluctuations of $\rho$ and $\Theta$ with respect to their QL duals:
\begin{equation} \Sigma = -\frac{1}{3}\,\left[\Theta-\Theta_q\right],\qquad
\EE = -\frac{\kappa}{6}\,\left[\rho-\rho_q\right].\label{SE2}\end{equation}
 
\section{Sufficient conditions for a positive back--reaction and an effective acceleration.}

Equation (\ref{efeacc}) provides the relation between back--reaction ($\QQ$) and the an effective acceleration mimicking dark matter. A necessary (but not sufficient) condition for the existence of this acceleration in a given comoving domain $\DD[r]$ of the form (\ref{domain}) is evidently 
\begin{equation} \frac{3}{2}\QQ=\langle \tilde\C(x,r)\rangle\geq 0,\qquad \tilde\C(x,r)\equiv \left(\Theta(x)-\Thetaav[r]\right)^2-\left(\Theta(x)-\Theta_q(x)\right)^2,\label{Qpos}\end{equation}
where we have used (\ref{SE2}) to eliminate $\Sigma^2$ in terms of $\Theta-\Theta_q$. Testing the fulfillment of this condition from the integral definitions (\ref{pmap}) and (\ref{QLmap}) is very difficult without resorting to numerical methods. However, for every domain $\DD[r]$ and every scalar we have $A(x)\geq 0\,\,\,\forall x\in\vartheta[r]\,\,\Rightarrow\,\, \Aav[r]\geq 0$ (though the converse is not necessarily true). Hence, a sufficient condition for the fulfillment of (\ref{Qpos}) in a given domain (\ref{domain}) is simply
\begin{equation} \tilde\C(x,r)\geq 0.\label{Cpos1}\end{equation}
Moreover, this condition is still too difficult to handle because of the dependence of $\tilde\C$ on points inside ($x<r$) and in the boundary ($x=r$) of the domain. Fortunately, by means of the following lemma we can find a condition equivalent to (\ref{Cpos1}) that depends only on the domain boundary and is applicable to any domain ($r$ variable).

\begin{quote}

\noindent \underline{Lemma 1}: $\langle \PP\rangle=0$ in every domain $\vartheta[r]$ for $\PP=\PP(x,r)$ given by
\begin{equation} \PP(x,r) = \left[\Theta(x)-\Thetaav[r]\right]^2-\left[\Theta(x)-\Theta_p(x)\right]^2,\label{newC}\end{equation}

\noindent \underline{Proof}.  Expanding (\ref{newC}) and applying (\ref{pmap}) and (\ref{pavemap}) we obtain with the help of (\ref{var})
\begin{equation} \langle\PP\rangle[r]=-\Thetaav[r]^2+\frac{1}{\VV_p(r)}\int_0^r{[2\Theta\Theta_p-\Theta_p^2]\,\VV_p'\,\dd x}.\end{equation}
Inserting $\Theta=\dot\VV_p'/\VV_p'$ and $\Theta_p=\dot\VV_p/\VV_p$ in the integrand above, and bearing in mind that $\Thetaav$ and $\Theta_p$ coincide at the domain boundary $x=r$, leads to the desired result:
\begin{equation} \langle\PP\rangle[r]=-\Thetaav^2[r]+\frac{1}{\VV_p(r)}\,\int_0^r{[\dot\VV_p^2/\VV_p]'\,\dd x}=-\Thetaav^2[r]+\Theta_p^2(r)=0.\end{equation}
An analogous result follows for the quasi--local average acting on a scalar like $\PP$ with $\langle\hskip 0.1 cm\rangle_q$ and $\Theta_q$ instead of $\langle\hskip 0.1 cm\rangle$ and $\Theta_p$.

\end{quote}

\noindent As a consequence of this lema, we have $\langle(\Theta(x)-\Thetaav[r])^2\rangle=\langle(\Theta(r)-\Theta_p(r))^2\rangle$ and so the sufficient condition for $\QQ\geq 0$ given by (\ref{Cpos1}) can be rewritten now as
\begin{equation}\C(r)\equiv \left(\Theta(r)-\Theta_p(r)\right)^2-\left(\Theta(r)-\Theta_q(r)\right)^2\geq 0,\label{Cpos2}\end{equation}
holding for comoving domains of the form (\ref{domain}). This condition is domain dependent, in the sense that it may hold for some domains and not for others. Since the condition for an effective acceleration in (\ref{efeacc}) is given as the average of the scalar $(2/3)\tilde\C (x,r)-(\kappa/2)\rho(x)$, lemma 1 also provides the following sufficient condition for its fulfillment 
\begin{equation}\ACal_(r) \equiv   \frac{2}{3}\C(r) - \frac{\kappa}{2}\rho(r)\geq 0.\label{Apos}\end{equation}
where $\C(r)$ is given by (\ref{Cpos2}). For the remaining of this paper we will examine conditions (\ref{Cpos2}) and (\ref{Apos}), finding out first the necessary restrictions on LTB models to comply with (\ref{Cpos2}), and then exploring for these models the fulfillment of (\ref{Apos}).  

\section{Probing the sign of the back--reaction term.}

In order to look at conditions (\ref{Cpos2}) and (\ref{Apos}), we need to explore the behavior of LTB scalars along radial rays of the hypersurfaces $\T$. For this purpose, the integral definitions (\ref{pmap}) and (\ref{QLmap}) yield the following properties of p--functions and averages and their quasi--local analogues: 
\ba
A'_p &=& \frac{3R'}{R}\,\frac{\FF_p}{\FF}\,\left[ A-A_p\right],\qquad\qquad\quad A'_q = \frac{3R'}{R}\,\left[ A-A_q\right]\label{prop1}\\
A(r)-A_p(r) &=& \frac{1}{\VV_p(r)}\int_0^r{A'(x)\,\VV_p(x)\,\dd x},\qquad A(r)-A_q(r) = \frac{1}{\VV_q(r)}\int_0^r{A'(x)\,\VV_q(x)\,\dd x},\label{prop2}   
\ea 
where $\FF_p$ is the p--function associated to $\FF(r)$ and $\VV_q=\int{\FF\,\dd\VV_p}=4\pi\int_0^r{R^2R'\dd x}=(4\pi/3)R^3(r)$. Considering (\ref{prop2}), condition (\ref{Cpos2}) for $\QQ\geq 0$ becomes
\begin{equation} \C(r) = \Phi(r)\,\Psi(r) \geq 0,\qquad\hbox{with:}\quad
\Phi(r) \equiv \int_0^r{\Theta'(x)\,\varphi(x,r)\,\dd x},\quad
\Psi(r) \equiv \int_0^r{\Theta'(x)\,\psi(x,r)\,\dd x},\label{Cpos3}\end{equation}
and with $\varphi$ and $\psi$ given by
\ba \varphi(x,r)=\frac{\VV_p(x)}{\VV_p(r)}-\frac{\VV_q(x)}{\VV_q(r)}=\frac{\VV_p(x)}{\VV_p(r)}\left[1-\frac{\FF_p(x)}{\FF_p(r)}\right],\label{phi_def}\\
\psi(x,r)=\frac{\VV_p(x)}{\VV_p(r)}+\frac{\VV_q(x)}{\VV_q(r)}=\frac{\VV_p(x)}{\VV_p(r)}\left[1+\frac{\FF_p(x)}{\FF_p(r)}\right],\label{psi_def}\ea
where we have used the relation $\VV_q/\VV_p=\FF_p$, which follows directly from (\ref{pmap}) and (\ref{QLmap}) and is valid for all domains. 

The fulfillment of (\ref{Cpos3}) clearly depends on the signs of $\varphi$ and $\psi$ (besides the sign of $\Theta'$) at all points in any given domain. This fulfillment might hold only in some domains and not in others. Since, by their definition, $\VV_p(0)=\VV_q(0)=0$ and $\FF(0)=1$, (\ref{psi_def}) and  (\ref{phi_def}) imply that for every domain we have $\psi(0,r)=0,\,\,\psi(r,r)=2$, whereas $\varphi(0,r)=\varphi(r,r)=0$. Thus, as long as $\FF$ and $R'$ are non--negative for all $x\in\vartheta[r]$, the sign of $\psi$ is non--negative, and so $\Psi$ basically depends only on the sign of $\Theta'$. On the other hand, the sign of $\varphi$ is not determined, and so the sign of $\Phi$ requires more examination as it depends on both: the sign of $\Theta'$ and the ratio $\FF_p(x)/\FF_p(r)$ (which relates to the sign of $\FF'$). Since $\FF$ and $R'$ can become negative in elliptic configurations where the $\T(t)$ have spherical topology, we will only consider in this paper ``open'' LTB models in which these slices are homeomorphic to $\textrm{\bf{R}}^3$. Since the sign of $\C$ depends on the sign of the product $\Phi\Psi$, we will need to obtain the conditions for both terms having the same sign.

In order to to probe condition (\ref{Cpos3}), we consider the following sign relations that emerge from (\ref{prop1})--(\ref{prop2}) and are valid for all $x\in\vartheta[r]$ in every domain and in every $\T$:
\ba \FF'\geq 0 \qquad \Rightarrow\qquad \FF\geq \FF_p\qquad \Rightarrow\qquad \FF(r)\geq \FF_p(x),\label{FFrpos}\\
\FF'\leq 0 \qquad \Rightarrow\qquad \FF\leq \FF_p\qquad \Rightarrow\qquad \FF(r)\leq \FF_p(x),\label{FFrneg}\ea
Considering (\ref{Cpos3})--(\ref{psi_def}) and the relations above, we have the following rigorous results:

\begin{quote}

\noindent \underline{Lemma 2}: Let $\Theta$ be a monotonous function in a domain (\ref{domain}) in a given $\T$, then :
\ba \QQ[r] &=& 0\qquad \hbox{if the domain belongs to a parabolic model},\label{L2p}\\
\QQ[r] &\geq& 0\qquad \hbox{if the domain belongs to a hyperbolic model},\label{L2a}\\
\QQ[r] &\leq& 0\qquad \hbox{if the domain belongs to an open elliptic model for which} \quad \FF'\leq 0,\label{L2b}\ea
The proof follows directly from (\ref{Cpos3})--(\ref{psi_def}). It is trivial for the parabolic case, since $\FF=1$ for every domain. For the hyperbolic and elliptic cases, having assumed a monotonous $\Theta$ implies that the sign of $\C$ only depends on the sign of $\varphi(x,r)$ in (\ref{phi_def}), but then regularity conditions \cite{ltbstuff,suss02} require $\FF'\geq 0$ to hold for all $r$ in hyperbolic models, while regularity admits (but does does not require) that $\FF'\leq 0$ holds for all $r$ in open elliptic models (we are not considering closed models, see \cite{sussBR} for the study of back--reaction in these models). Hence, we have $\C\geq 0$ and $\C\leq 0$, respectively,  for the hyperbolic and elliptic cases, and the result follows. \\     

\end{quote}

\noindent
Notice that (\ref{L2p}) is domain independent (holds at every domain $\vartheta[r]$), whereas (\ref{L2a}) necessarily becomes domain independent in those $\T$ of hyperbolic models for which $\Theta$ is monotonous for all $r$. For hypersurfaces $\T$ of open elliptic models in which $\Theta$ is monotonous for all $r$, then (\ref{L2a}) may become domain dependent: it holds for all domains sufficiently close to the center (where $\FF'\leq 0$) but may not hold if $\FF'$ changes sign, though (\ref{L2a}) is domain independent if $\FF'$ does not change sign (see \cite{sussBR}).

The restriction that $\Theta$ be monotonous for all $r$ does not hold, in general, for all $\T$ in any hyperbolic or open elliptic model (see \cite{sussinprep} for details). However, the effect of $\Theta'$ changing sign simply makes the sign of $\QQ$ domain dependent along any given $\T$. The same remark holds for open elliptic models for which $\FF'$ changes sign:
      
\begin{quote}

\noindent \underline{Lemma 3}: Consider the following two possible configurations along a given $\T$: (i) $\Theta'$ changes sign at a given $x=y\in\vartheta[r]$ in a hyperbolic model, and (ii) $\FF'$ and $\Theta'$ respectively change sign at a $x=y_1,\,x=y_2$ inside $\vartheta[r]$ in an open elliptic model. For both cases there are always domains $\vartheta[r_0]$ such that $\QQ[r]\geq 0$ holds for all $r>r_0$.\\

The proof in both cases is based on the fact that the zero of $\Theta'$ is fixed at any given $\T$, whereas the choice of domain is arbitrary. Consider the hyperbolic case: the zero of $\Theta'$ implies that $\Theta$ is monotonous in the full ``external'' range $r>x>y$. Hence if regularity conditions hold, $r$ can take arbitrarily large values and we can always find sufficiently large domains such that $r\gg y$ holds, and thus we can use (\ref{FFrpos}) to obtain the expected sign in the integration of $\Phi$ and $\Psi$ in (\ref{Cpos3}) along the range $y<x\leq r$. For sufficiently large domains the ``external'' contribution $y<x\leq r$ outweighs that of the inner range $0\leq x\leq y$. The same situation occurs in the elliptic case, for which $\FF'$ must pass from negative to positive at $x=y_1$. The zeroes of $\Theta'$ and $\FF'$ imply that $F$ and $\Theta$ are monotonous for $x>\hbox{max}(y_1,y_2)$, and thus the same argument applies. The reader is advised to consult \cite{sussBR} for further detail on these proofs. 
    
\end{quote}

\noindent
As a consequence of lemmas 2 and 3, the condition $\QQ>0$ is compatible with both, negative and positive spatial curvature (hyperbolic and open elliptic models),  at least in their asymptotic radial range. We need now to find if the positive back--reaction is comparable to the average density. 
\begin{figure}[htbp]
\includegraphics[width=4in]{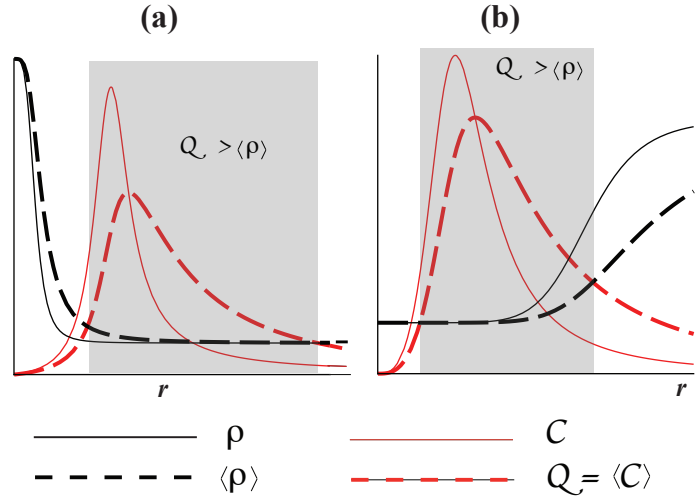}
\caption{{\bf Scenarios with positive effective acceleration.} The figure displays, for a density clump (a) and a void (b), the radial profiles of $\rho(x)$ (black solid curve) and $\C(x)$ (red solid curve), compared with $\rhoav$ (black dotted curve) and back--reaction $\QQ\propto \langle\C\rangle$ (red dotted curve) along a given regular hypersurface $\T$ of either a hyperbolic or an open elliptic model. Notice an intermediate region (gray area) where density is low and nearly homogeneous, while the growth of back--reaction is driven by the gradients of spatial curvature and reaches a maximum, thus allowing for (\ref{Apos}) to hold.}
\label{fig2}
\end{figure} 
\section{Probing the sign of the effective acceleration.}

In order to examine (\ref{Apos}), we will need the following rigorous results concerning the radial profiles of scalars $A$ along the $\T$:

\begin{quote}

\noindent \underline{Lemma 4}. If $R'>0$ everywhere and $A(r)\to A_0=$ constant as $r\to\infty$, then $A_p(r)\to A_0$ and $A_q(r)\to A_0$ as $r\to\infty$, while $A_p'(r)\to 0$ and $A_q'(r)\to 0$ in this limit.\\

\noindent \underline{Proof}. We consider the case when $A'\geq 0$ for sufficiently large values of $r$. The case when $A'\leq 0$ is analogous. If the limit of $A$ as $r\to\infty$ is $A_0$, then for all $\epsilon>0$ there exists $L(\epsilon)$ such that $A_0-\epsilon<A(r)<A_0$ holds for all $r>L(\epsilon)$. Constraining $A$ by means of this inequality in the definitions of $A_p$ and $A_q$ in (\ref{pmap}) and (\ref{QLmap}) leads immediately to $A_0-\epsilon<A_q(r)<A_0$ and $A_0-\epsilon<A_p(r)<A_0$. Hence, the limit of $A_p$ and $A_q$ as $r\to\infty$ is also $A_0$. The limits $A_p'(r)\to 0$ and $A_q'(r)\to 0$ follow trivially.\\

\noindent \underline{Lemma 5}. If there is a zero of $A'$ at $x=y$ and $R'>0$ in $\vartheta[r]$, then for sufficiently large $r$ there will be a zero of $A'_p$ at $x=r_1>y$ and a zero of $A'_q$ at $x=r_2>y$, with $A(r_1)=A_p(r_1)$ and $A(r_2)=A_q(r_2)$.\\ 

\noindent \underline{Proof.} Let $A'$ pass from positive to negative at $x=y$. As $x$ reaches $y$ the first integral in (\ref{prop2}) is still positive and so $A(y)>A_p(y)$, but for $y<x<r$ the integrand becomes negative, and so the contributions to the integral are increasingly negative. Since $\VV_p(x)/\VV_p(r)$ is increasing,  if $r$ is sufficiently large, then a value $x=r_1>y$ is necessarily reached so that this integral vanishes (thus $A(r_1)=A_p(r_1)$). From (\ref{prop1}), we have $A'_p(r_1)=0$ and $A'_p<0$ for $x>r_1$. An analogous situation occurs when $A'$ passes from negative to positive. The proof is identical for $A_q$, but using the second integral in (\ref{prop2}).

\end{quote}

\noindent
In order to apply these lemmas to the case $A=\C$ with $\C$ given by (\ref{Cpos2}), we use (\ref{prop1}) to rewrite this scalar in terms of the gradients of $\Theta_q$ and $\Theta_p$ as
\begin{equation} \C = \left(\frac{R}{3R'}\right)^2\,\left[\left(\Theta_p'\frac{\FF}{\FF_p}\right)^2-\Theta_q'{}^2\right]=\left(\frac{R}{3R'}\right)^2\,\left[\Theta_p'\frac{\FF}{\FF_p}-\Theta_q'\right]\left[\Theta_p'\frac{\FF}{\FF_p}+\Theta_q'\right].\label{CCgrads}\end{equation}
Considering now LTB models (hyperbolic and open elliptic) in which the scalars $\{\Theta,\rho,\RR\}$ tend to nonzero finite values $\{\Theta_0,\rho_0,\RR_0\}$ as $r\to\infty$ (an asymptotic FLRW state in the radial direction), then Lemma 4 and (\ref{CCgrads}) imply that $\C\to 0$,\,\,$\Theta-\Theta_p\to 0$ and $\Theta-\Theta_q\to 0$ as $r\to\infty$, and also $\Theta'_p$ and $\Theta'_q$ vanish in this limit. Since $\C(0)=0$ follows from (\ref{prop2}), then domains must exist for which $\C\,'$ must have a zero for some $x=y\in\vartheta[r]$. Lemma 5 implies then that domains must exist in which $\QQ\,'=(2/3)\langle\C\rangle\,'$ has also a zero at $r_1>y$, corresponding to a local maximum of $\QQ$ where it reaches its maximal value in the domain (see figure 2). 

The next step is to compare $\C$ and $\QQ$ with $\rho$ and $\rhoav$ in order to test the fulfillment of (\ref{Apos}). For this purpose we note that $\Theta,\,\Theta_p,\,\Theta_q$ are respectively related to $\rho,\,\rho_p,\,\rho_q$ and $\RR,\,\RR_p,\,\RR_q$ by the constraints (\ref{cHam_13}), (\ref{ave_Fried}) and (\ref{QLfried}). Hence, the magnitude of the expansion gradients is closely connected with the magnitude of the gradients of the density and spatial curvature. Given the fact that $\QQ(0)=0$ and $\QQ\to 0$ as $r\to\infty$, reaching a maximum in the intermediate range, the best possible situation in which (\ref{Apos}) could hold is if $\QQ$ reaches its maximum in the same region where $\rho$ (and thus $\rhoav$) has a low and almost constant value. In this scenario we would have a large intermediate region in which $\rho(r)/\rho(0)\ll 1$ and $\rho'\approx 0$ (and thus $\rhoav[r]/\rho(0)\ll 1$ and $\rhoav'\approx 0$ hold), so that the growth of $\C$ (and thus $\QQ$) is driven by the gradients of the spatial curvature. The scenario described above is illustrated in figure 2 for a density clump and void profile.

\section{Conclusion.} 

We have introduced quasi--local variables and averages in LTB dust models. These scalar variables are very useful to discuss various theoretical issues concerning these models, such as the application of Buchert's scalar averaging formalism~\cite{buchert,ave_review}. We have found in this paper analytic conditions for a positive back--reaction term $\QQ$ and for an effective acceleration mimicking dark energy in these models. The present paper provides a quick summary of comprehensive articles \cite{sussQL,sussBR,sussinprep} that are currently under revision, all of them dealing with different aspects of radial profiles of scalars and the issue of back--reaction in LTB models. We have chosen here the simplest boundary conditions (radial asymptotical homogeneity) to illustrate how generic LTB configurations with open topology (hyperbolic and elliptic) can fulfill the conditions for such an effective acceleration. More general conditions are examined in \cite{sussinprep}. It is very likely that the astrophysical and cosmological effects of Buchert's formalism will require numerical methods applied to more ``realistic'' configurations not restricted by spherical symmetry and compatible with observations. The analytic study carried on here and in the associated references can certainly provide a useful guideline for this important task.


\begin{theacknowledgments}
  The author acknowledges financial support form grant PAPIIT--DGAPA IN--119309. 
\end{theacknowledgments}



\bibliographystyle{aipproc}   

\bibliography{sample}

\IfFileExists{\jobname.bbl}{}
 {\typeout{}
  \typeout{******************************************}
  \typeout{** Please run "bibtex \jobname" to optain}
  \typeout{** the bibliography and then re-run LaTeX}
  \typeout{** twice to fix the references!}
  \typeout{******************************************}
  \typeout{}
 }

\end{document}